\documentclass[aps,pre,twocolumn,groupedaddress,showpacs]{revtex4}

\usepackage{amsmath,amssymb,amsfonts}
\usepackage{graphicx}
\usepackage{color}
\begin{document}
\title{Network Evolution Based on Centrality}
\author{Michael D. K{\"o}nig}\affiliation{Chair of Systems Design, D-MTEC,
ETH Zurich, CH-8032 Zurich, Switzerland,}
 \affiliation{Department of Economics, Stanford University, CA-94305-6072, United States of America.}
\author{Claudio J. Tessone}
\affiliation{Chair of Systems Design, D-MTEC,
ETH Zurich, CH-8032 Zurich, Switzerland.}
\date{\today}
\begin{abstract}
  We study the evolution of networks when the creation and decay of links are based on the position of nodes in the network measured by their centrality. 
  We show that the same network dynamics arises under
  various centrality measures, and solve analytically the network
  evolution. During the complete evolution, the network is
  characterized by nestedness: the neighborhood of a node is
  contained in the neighborhood of the nodes with larger degree. We
  find a discontinuous transition in the network density between hierarchical and homogeneous
  networks, depending on the rate of link decay. We also show that this evolution
  mechanism leads to double power-law degree distributions, with
  interrelated exponents.
\end{abstract}
\maketitle

\section{Introduction}
The underlying mechanisms of link formation governing the evolution of
a network ultimately determine its emergent properties at the
aggregate level \citep{albert02:_statis, Dorogovtsev:2006}. In
particular, there exists ample empirical evidence that the network
evolution can be driven by {\em centrality}, where nodes with
higher centrality are more likely to form or receive links 
\citep{nerkar05:_evolut_of_r_d_capab,
  gulati99:_where_do_inter_networ_come_from}. The
notion of centrality was recognized to play a fundamental role
in the most despair fields, ranging from dynamical systems
\cite{restrepo05:_onset_of_synch_in_large}, synchronization
\cite{Nishikawa:2003}, biology
\cite{jain98:_autoc_sets_growt_compl_evolut_model}, and economics
\cite{nerkar05:_evolut_of_r_d_capab, gulati99:_where_do_inter_networ_come_from}.
In spite of its importance, a formal understanding of how networks evolve when
the formation of links depends on the centrality of the nodes 
is still missing.

Depending on the context, several measures of centrality have been
introduced to quantify the importance of the position of a node in a network:
\textit{degree}, \textit{eigenvector}, \textit{betweenness},
\textit{closeness}, \textit{PageRank} and \textit{Bonacich centrality}
are the most prominent ones \citep{wasserman94:_social_networ_analy,
  brin98:_anatom_of_large_scale_hyper}. Due to this variety, few
attempts have been made so far to elucidate the common features underlying
the emergent properties of networks evolving by centrality
\cite{holme06:_dynam_networ_agent_compet_high}.

At the macroscopic level,
some real world networks exhibit a high degree of clustering while,
coincidentally, their degree distributions show power-law tails. Taken
together, these two characteristics indicate a hierarchical
organization in the network
\citep{ravasz03:_hierar_organ_in_compl_networ}. In social and economic
\citep{may08:_ecolog_for_banker, AAkerman2010}, as well as biological systems
\citep{bastolla09:_archit_of_mutual_networ_minim}, it has been found
that the hierarchical organization of networks can further be
characterized by {\em nestedness}
\cite{saavedra08:_simpl_model_of_bipar_cooper,
  bastolla09:_archit_of_mutual_networ_minim}: the neighborhood of a
node is contained in the neighborhood of the nodes with higher
degrees. In these examples, the extent of nestedness (defined as the
fraction of links belonging to the nested structure) was shown to be
above $93 \%$ \cite{saavedra08:_simpl_model_of_bipar_cooper}. A recent study  \cite{Leskovec2009} also finds nested core-periphery structures in over $100$ large sparse real-world social and information networks. 

In this paper, we study a model of network evolution where links are
created or removed based on the centrality of the nodes incident to
the links \cite{koenig09:_social_networ_growt_agent_compet_high_centr}.  We show that in this model the network evolution is independent of the particular
centrality measure used. Thus, for the first time, this model provides
a general framework to study the evolution of networks under various
measures of centrality.  We show that there exist stationary
networks which are highly hierarchical when the rate of link creation
is low. Moreover, the networks are {\em nested} during the complete
network evolution.  As we show, both, a hierarchical organization as well as
network nestedness can be the outcomes of a centrality based network
formation process. We further show that in this framework, double
power-law degree distributions
\citep{csanyi04:_struc_of_large_social_networ,
  boss04:_networ_topol_of_inter_market, gjoka10_walkingfb} can be stationary solutions,
and that each power-law exponent has a univocal relation to the 
other. 

The paper is organized as follows. In Section \ref{sec:model} we introduce the basic network formation process and discuss the generality of its underlying assumptions, by showing the independence of the dynamics with respect to the particular centrality measure used. Next, in Section \ref{sec:results} we provide the characterization of the asymptotic network structures generated by our network formation process. In Section \ref{sec:generalized} we then extend the basic network formation process by allowing for heterogeneous linking probabilities among the nodes in the network and study the effect this has on the emerging network structures. Finally, Section \ref{sec:conclusions} concludes.

\section{Model studied} 
\label{sec:model}
We consider a network composed of $N$ nodes, initially
connected by an arbitrary network. Each node has a centrality
associated to it. 
At a constant rate (set arbitrarily to one and {\em a priori} equal for all nodes) a node is randomly selected and modifies its neighborhood: with probability $\alpha \in [0,1]$, it creates a link to the node with the highest centrality it is not already connected to. With the complementary probability $1-\alpha$, a link of the selected node decays. If this happens then the node removes the link to the neighbor with the lowest centrality. If the node is
connected to all the other nodes in the network (resp.~it is
isolated), and it has to create (resp.~remove) a link, nothing
happens. One can show that the
network formation process is ergodic, and starting from any initial
network yields the same asymptotic results. Thus, and without any loss of
generality, in the following we consider an empty network as
initial condition.

Note that, when links are created, we could assume that a node has only local information of the network \cite{friedkin83:_horiz_of_obser_and_limit} and creates a link to the one with the highest
centrality in its second-order neighborhood. It turns out that this leads to the same network evolution process. This makes sense in situations where centrality
is known {\em ex ante}, for example, when centrality is a measure of
performance in inter-organizational networks
\citep{mehra01:_social_networ_of_high_and}, or it indicates the fitness
of biological species 
\citep{jain98:_autoc_sets_growt_compl_evolut_model}.

The general dynamics we have introduced above can be applied to different areas. In the following we provide a few illustrative examples.

A first example can be given for ecological networks. Consider a
population of biological species in a catalytic network
\citep{jain98:_autoc_sets_growt_compl_evolut_model}. Let species
$i=1,\dots,N$ have a fitness values $y_i \ge 0$ that evolve according
to the dynamics 
$$
\dot{y}_i=\sum_{j=1}^N a_{ij} y_j - \phi y_i, \phi \ge 0,
$$
where $a_{ij} \in \{0,1\}$ is the $ij$-th element of the
symmetric adjacency matrix $\mathbf{A}$. In terms of relative fitness
$x_i = y_i/\sum_{j=1}^N y_j$, we get 
$$
\dot{x}_i = \sum_{j=1}^N a_{ij} x_j - x_i \sum_{k,j=1}^N a_{kj} x_j.$$
This dynamics has a fixed
point given by the eigenvector $\mathbf{v} \ge 0$ corresponding to the
largest real (Perron-Frobenius) eigenvalue $\lambda_{\text{PF}}$ of
$\mathbf{A}$. Hence, in this model, the stationary fitness distribtion is directly given by the
eigenvector centrality. Applying this centrality measure to our network formation process mimics an evolutionary process in which links to high fitness species are created while links to low fitness species decay.

A second example comes from a socio-economic context. Consider a population of agents whose payoffs are interdependent in a network. The agents choose a contribution level $x_i \ge 0 $ and
receive a payoff $\pi_i$ given by 
$$
\pi _{i} = x_{i}- \frac{1}{2}x_{i}^{2}+\lambda
\sum_{j=1}^{N}a_{ij}x_{i}x_{j},
$$ 
where $\lambda < 1/\lambda_{\text{PF}}$
\citep{ballester06:_who_networ}. Then, the unique Nash equilibrium satisfying the first-order condition $\partial \pi_i / \partial x_i=0$ is
given by the Bonacich centrality \citep{wasserman94:_social_networ_analy},
$$
\mathbf{x}^* = (\mathbf{I}-\lambda \mathbf{A})^{-1}\mathbf{1}.
$$ 
The linking dynamics introduced above corresponds to a game in which agents form links that maximize their Nash equilibrium payoffs $\pi_i(\mathbf{x}^*)$ in each period \citep{koenig09:_social_networ_growt_agent_compet_high_centr}. 

Further examples include degree centrality, closeness centrality
\citep{holme06:_dynam_networ_agent_compet_high}, betweenness
centrality \citep{wasserman94:_social_networ_analy}, PageRank
\citep{brin98:_anatom_of_large_scale_hyper} and random walk
  centrality \citep{noh04:_random_walks_compl_networ}.  
One can also show that the links created (removed) in our model are the ones which increase the most (decrease the least) the largest eigenvalue $\lambda_{\text{PF}}$. These links were shown to modify the most the dynamical properties of the system \citep{restrepo06:_charac_dynam_impor_of_networ}.

\begin{figure}
\begin{center}
\includegraphics[angle=0,width=0.5\linewidth]{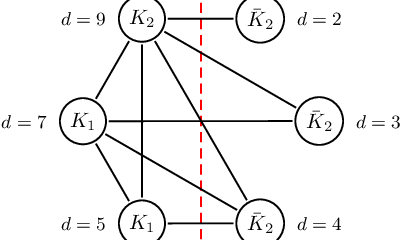}
\includegraphics[angle=0,width=0.4\linewidth]{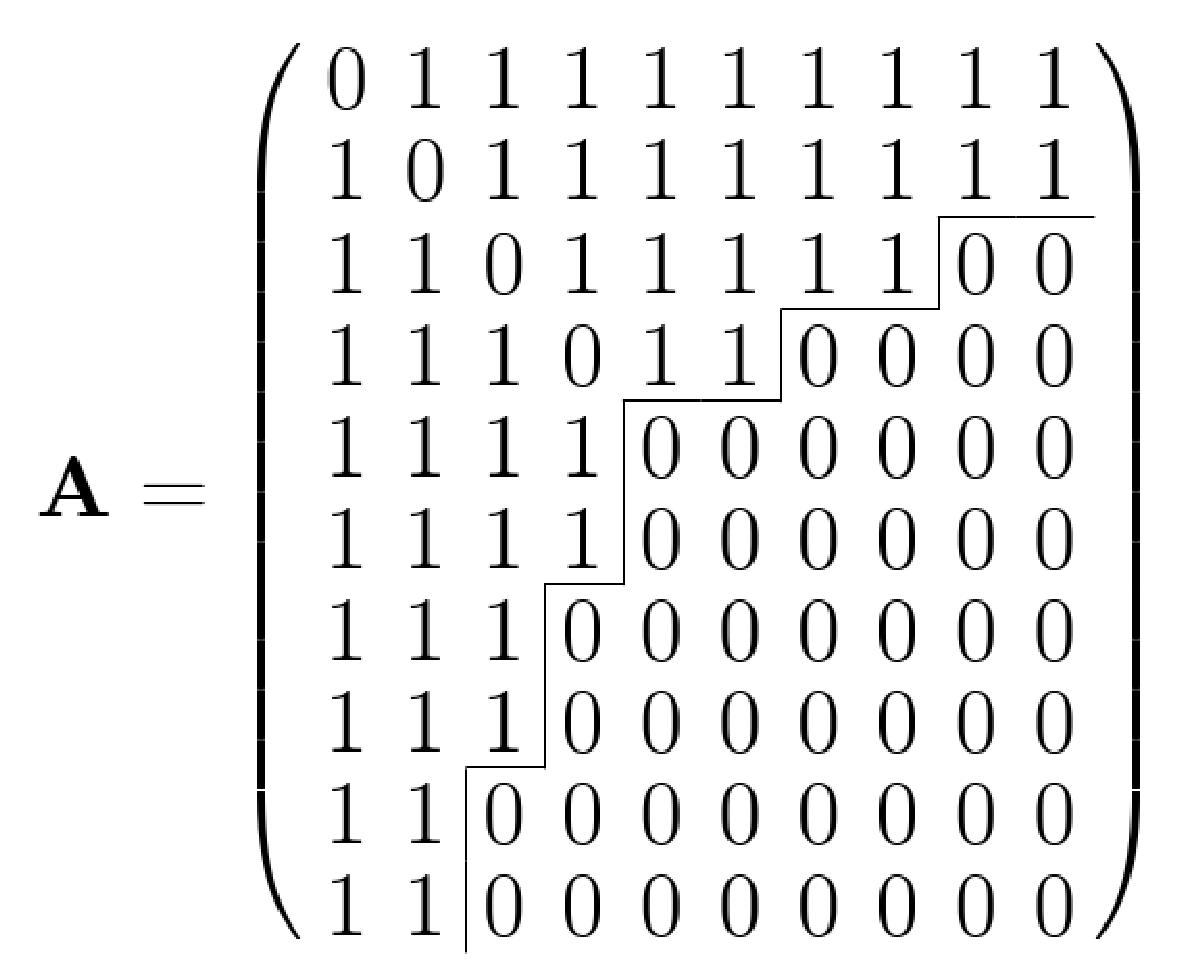}
\end{center}
\caption{Representation of a nested network (left) and the associated
  stepwise adjacency matrix (right) with $N=10$ nodes. A nested
  network can be partitioned into subsets of nodes with the same
  degree (each subset is represented by circle, next to which the
  degree $d$ of the nodes in the subset is indicated). A line
  connecting two subsets indicates that there exists a link between
  each node in one set to all nodes in the other set. The union
  of the sets represented by the circles to the left of the dashed
  line induce a dominating set, while to the right the circles
  indicate independent sets. In the matrix $\mathbf{A}$ to the right,
  the zero-entries are separated from the one-entries by a
  step-function, $h(r)$, of the rank degree of the nodes.}
\label{fig:nested-split-graph}
\end{figure}

The distinctive characteristic of our network formation process that allows us to incorporate various centrality measures is the fact that, at every time step $t=0,1,2,\dots$, our dynamics yields a network whose adjacency
matrix $\mathbf{A}$ is \textit{stepwise}: the nodes can be ordered by
their degree, such that the zero/one entries in the adjacency matrix
are separated by a monotonic step-function $h(x)$ (see
Fig.~\ref{fig:nested-split-graph}, right), where $x=1-r$, and $r$ is the
degree rank of a node. 
Networks with a stepwise adjacency matrix are also known as \textit{threshold
  networks} \citep{mahadev95:_thres_graph_and_relat_topic,
  hagberg06:_desig_thres_networ_with_given}.
In such a network, if two nodes $i$ and $j$ have degrees such that $d_i <
d_j$, then their neighborhoods satisfy $N_i \subset N_j$. Thus, these
networks are characterized by nestedness.
Moreover, the nodes can be partitioned into a
\textit{dominating set} and \textit{independent sets}. In the dominating
set $S$, every node not in $S$ is linked to at least one member of
$S$. Conversely, an independent set is one in which no two nodes are
adjacent (see Fig.~\ref{fig:nested-split-graph}, left).  

We now prove by induction that the adjacency matrix $\mathbf{A}$ representing the
state of the network at every time step is stepwise for the case of
eigenvector centrality. First, at time $t=0$, the first link added
generates a (trivial) stepwise matrix.  Next, let us assume that this
is true at time $t \ge 0$. Consider the creation of a link $ij$.  Then
$$
v_i = \frac{1}{\lambda_{\text{PF}}} \sum_{k=1}^n a_{ik} v_k =
\frac{1}{\lambda_{\text{PF}}} \sum_{k \in N_i} v_k.$$
Thus, the larger is the
degree of a node $i$, the higher is its eigenvector component
$v_i$. In this way, the eigenvector centrality of the nodes is ranked
in the same way as their degree. Therefore, for the model studied, a
node has to establish a link to a node with the highest degree it is
not connected to.  This preserves the stepwise property of
$\mathbf{A}$ (see Fig.~\ref{fig:nested-split-graph}, right).
Similarly, for the removal of links, the node with the lowest degree
among the neighbors is the least central one, and removing a link to
it preserves the stepwise property of $\mathbf{A}$.

The nested neighborhood structure allows us to use similar arguments
for other centrality measures. Consider two nodes $i$ and $j$ in a
nested graph with $d_i>d_j$. All walks starting at node $j$ are
contained in the set of all walks starting from node $i$ (after
exchanging the starting node $j$ with $i$).  This implies that $i$ has
a higher centrality than $j$ for any centrality measure that is based
on walks or paths in the network. Hence, a proof by induction shows
that the ranking of nodes by degree is equivalent to the ranking by
centrality for this family of centrality measures. In general, this
dynamics leads to a self-reinforcement of the nested structure.

\section{Results}
\label{sec:results}

Given the symmetry of the adjacency matrix $\mathbf{A}$, in order to
solve the dynamic evolution of the network, it is enough to solve the
dynamics for the nodes belonging to the independent sets (see
Fig.~\ref{fig:nested-split-graph}).  Let us denote by $n(d,t)$ the
number of nodes in the independent sets with degree $d$ at time
$t$.  The dynamic evolution of these populations can be 
written as a rate equation,
\begin{eqnarray}
  \partial_t n(d,t) 
  & = &   \omega[d+1 \to d ] \, n(d+1,t) + \label{eq:disc:intermediate}\\
  &   &  \omega[d-1 \to d) \, n(d-1,t) - \nonumber \\
  & &  \left( \omega[d \to d-1] + \omega[d \to d+1] \right) \, n(d,t),\nonumber   
\end{eqnarray}
where the transition rates are simply $\omega[d \to d+1] = \alpha /
N$, $\omega[d \to d-1] = (1-\alpha) / N$.  In
Eq.~(\ref{eq:disc:intermediate}) we have neglected the contributions
of the nodes in the corresponding dominating set (which are selected with
a probability $\sim \mathcal{O}(N^{-1})$) in the dynamics of the nodes in the
independent sets.
The dynamics studied is restricted to the
profile separating (non-)existing edges, and is thus related to 
surface-growth models, such as those of polynuclear growth; because of this, it can
also be linked to the one-dimensional Ising model with Kawasaki dynamics
\cite{hinrichsen2000}. 

The dynamic evolution of the network can be written in terms of its
degree distribution $P(k; t)=n(d,t)/N$, where $k=d/N$ denotes the
normalized degree.  For a finite population, the minimum increment
possible in degree is $\delta k = 1/N$. At leading order in $\delta
k$, the dynamic evolution corresponding to
Eq. (\ref{eq:disc:intermediate}) is given by
\begin{equation}
\begin{split}
  \partial_t P(k; t) &
  = (1-2 \alpha)\, \partial_k P(k; t) \\
  &  \qquad \qquad + \delta k \, \partial^2_{kk} P(k; t) + {\mathcal{O}(\delta k^2)}, \\
  \partial_t P (0;t) &
  = (1-2\alpha) \left( \delta k + \partial_k P(0;t ) \right. \\
  & \left. \qquad \qquad - \alpha \, \delta k\, P(0;t ) \right) + {\mathcal{O}(\delta k^2)},\\
\end{split}
\label{eq:cont:intermediate:full}
\end{equation}
with the additional boundary condition $P (1,t)={\mathcal{O}(\delta
  k^2)}$ and an initial condition $P(k,0) = \delta(k)$.

When the terms of order $\delta k$ can be neglected,
Eq.~(\ref{eq:cont:intermediate:full}) becomes a usual drift equation
whose stationary solution is either a complete network for
$\alpha>1/2$ (when the link decay is low), or empty for
$\alpha<1/2$. The reason for this lays in the change of sign in the drift coefficient in such equation.
Thus, there is a discontinuous phase transition in terms of the network density as a function of the decay rate $\alpha$.
If $\alpha$ is small (and the link decay is high), the
network rapidly converges to a hierarchical structure, where only a few
nodes immediately become central, and they remain in this central
position during the network evolution. In this case it is the
competition driven dynamics for centrality which leads to the
spontaneous emergence of hubs
\citep{anghel04:_compet_driven_networ_dynam}.

There exists a first order phase transition in the network density
that gives rise to nontrivial effects around the critical point
$\alpha = 1/2$.  If $ | 1 - 2 \alpha | / \delta k \sim {\mathcal
  O}(1)$, then the diffusion term in
Eq.~(\ref{eq:cont:intermediate:full}) is not negligible anymore. Time
scales must be rescaled to $\tau \equiv t \, \delta k$, and  we
get the Fokker-Planck equations
\begin{eqnarray}
\partial_\tau P(k; \tau) &=& (1-2 \alpha) \,  \partial_k  P(k; \tau) + \partial^2_{kk} P(k; \tau) \\
  \partial_\tau P (0; \tau) &   = & \frac{1-2\alpha}{\delta k}  \partial_k P(0; \tau )  .
\end{eqnarray}
This prescription allows to relate the width of the transition from
sparse to dense networks: it must be that $|1-2\alpha| \sim {\mathcal
  O}(1)$, or conversely, $\Delta \alpha \sim N^{-1}$.

We now study the stationary solutions for all values of $\alpha \in
[0,1]$. First, notice that the network obtained for a value of
$\alpha>1/2$ is the complement of the network obtained for
$1-\alpha<1/2$. Thus, in the following we consider only values of
$\alpha \le 1/2$. The step-function $h(x)$ can be decomposed in a
part $h_u(x)$ below the diagonal and a part $h_l(x)$ above the
diagonal of $\mathbf{A}$ (see Fig.~\ref{fig:nested-split-graph}, right
panel). The point $x^*$ is implicitly defined by $ h_u(x^*) =
h_l(x^*)$, where the step-function $h(x)$ intersects with the
diagonal. 
Let $P(k) $ denote the stationary degree
distribution. We have that $h_u(x) = \int_0^{1-x} P(k) dk$. From the
stationary solution of Eq.~(\ref{eq:cont:intermediate:full}) we find 
$$
h_u(x) =
\mathcal{N} e^{-2(1-2 \alpha) x}, 
$$
with
$$\mathcal{N} =\frac{2(1-2\alpha)}{1-e^{- 2(1-2\alpha) N}}.
$$
This result for the functional form of the step-function is valid for the
elements below the diagonal, i.e.~for the nodes with low degree. 

\begin{figure}
\begin{center}
  \includegraphics[width=0.8\linewidth]{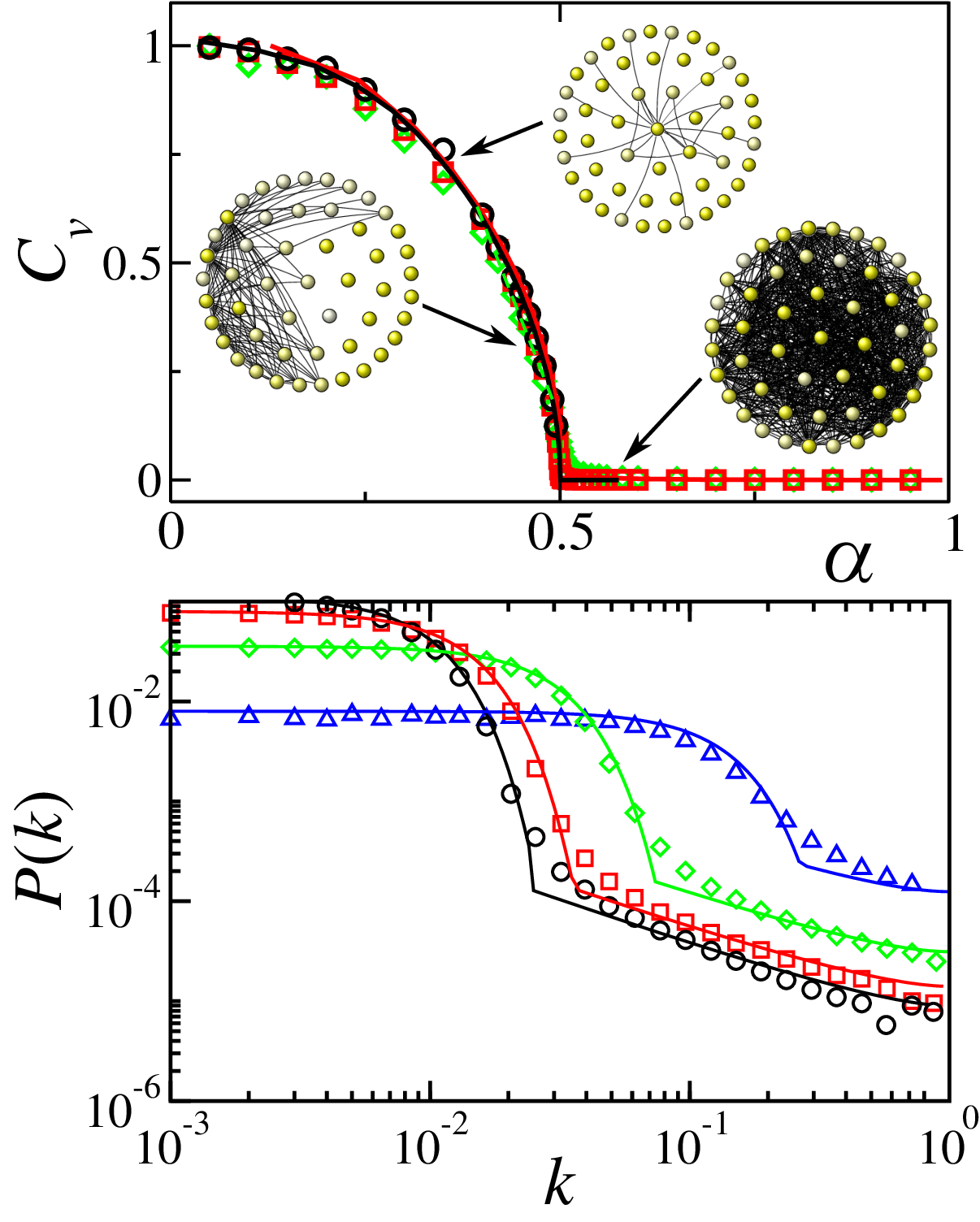} 
\end{center}
\caption{(Upper panel) Eigenvector centralization $\mathcal{C}_{v}$ in
  stationary networks as a function of the link formation probability
  $\alpha$ for different system sizes $N=100$ ({\large $\circ$}),
  $N=1000$ ({\small $\color{red} \Box$}) and $N=5000$ ({$\color{green}
    \Diamond$}). Results of numerical simulations are superimposed
  with lines representing the analytical prediction.  (Lower panel) Degree
  distributions of stationary networks for different values of $\alpha
  = 0.45$ ({\large $\circ$}), $0.48$ ({$\color{red} \Box$}), $0.49$
  ({$\color{green} \Diamond$}), $0.495$ ({$\color{blue}
    \triangle$}) and system size $N=5000$.  The figure reveals that the
  leading part of the distribution is exponential, while a logarithmic
  binning shows a power-law tail with exponent $-1$.}
\label{fig:density-centrality-network-examples}
\end{figure}

We
now turn our attention to the high degree, central nodes. From the 
symmetry of the adjacency matrix, one finds that $h_l(x)$ for these
nodes satisfies $x = \mathcal{N} e^{-2 (1-2 \alpha ) h_l(x)}$. Thus, inverting this expression we get 
$$
h_l(x) = \frac{\ln (\mathcal{N}) - \ln
(x)}{2(1-2 \alpha )}.
$$
  Conversely, the degree distribution is given
by $P(k) = -h'(1-k)$, from which the following stationary degree
distribution is found
\begin{equation}
P(k) =
\begin{cases}
  {\mathcal N} e^{-2(1-2\alpha)k},  & \text{if } k<1-x^*,\\
  \frac 1 { 2 (1-2\alpha)} k^{-1}, & \text{if } k>1-x^*.
\label{eq:basic-model-stationary-degree-distribution}
\end{cases}
\end{equation}
In particular, for $\alpha=1/2$, it results in a uniform distribution
$P(k)=1/N$.  Degree distributions for different values of $\alpha$ in
the stationary state can be seen in
Fig.~\ref{fig:density-centrality-network-examples}. 

In these nested structures, the adjacency matrix is completely
determined by the corresponding degree distribution from
Eq.~(\ref{eq:basic-model-stationary-degree-distribution}) or, conversely, from
the profile function $h(x)$. Thus, it is
possible to compute any network statistic of interest when the degree
distribution is known.  In doing so, one can show that the stationary
networks emerging in the link formation process are characterized by
short path length, high clustering, 
negative degree-clustering correlations and dissortativity.

The emerging networks also show a clear core-periphery structure, which
can be measured by their centralization. To quantify this, 
we compute the degree of centralization of the network $\mathcal{C}_v$, as
\citep{wasserman94:_social_networ_analy}
\begin{equation}
\mathcal{C}_v = \frac{\sum_i \left( \mathcal{C}_v(i^*)-\mathcal{C}_v(i)
\right)} { \sum_{j } \left( \mathcal{C}^*_v%
(j^*)-\mathcal{C}^*_v(j) \right)},
\end{equation}
where ${\mathcal C}_v(i)$  if the eigenvector centrality of node $i$, 
 $i^*$ is the node with the largest centrality in the network. The denominator normalizes the value between zero and one, by the computation of the centralization $\sum_{j } \left( \mathcal{C}^*_v%
(j^*)-\mathcal{C}^*_v(j) \right)$ of a star network  the the same maximum degree as the considered one.
In Fig.~\ref{fig:density-centrality-network-examples} (upper panel) we show the transition from hierarchical to decentralized networks measured in terms of the degree of centralization of the network, as a function of the parameter $\alpha$. In the same plot, also exemplary stationary networks are depicted.
It can be seen that there
 exists a transition at $\alpha=1/2$ from highly centralized to highly
 decentralized networks. 
This means that for low arrival rates of
linking opportunities $\alpha$ (and a strong link decay) the
stationary network is strongly centralized, while for high arrival
rates of linking opportunities, stationary networks are dense and
largely homogeneous.

\begin{figure}[t]
\begin{center}
 \includegraphics[width=0.8\linewidth]{4.eps}
\end{center}
\caption{(Upper panel) Degree distribution for different exponents in the
  head of the distribution $\eta=1.2$ ({$\color{black} \nabla$}),
  $1.5$ ({$\color{red} \triangle$}), $2$ ({$\color{green} \Diamond$}),
  $2.5$ ({$\color{blue} \Box$}), $3$ ({\large $\color{magenta}
    \circ$}). If a nested network exhibits a power-law in the head
  (tail) of the degree distribution then the distribution will also
  exhibit a power-law behavior in the tail (head), with an exponent
  that can be completely determined by the head (tail). (Lower panel)
  Power-law exponents for the tail of the degree distribution,
  i.e.~$k\to\infty$ ({\small $\color{red} \Box$}), and the head of the
  distribution, i.e.~$k\to 0$ ({\large $\circ$}), as a function of the
  power-law exponent of the head. The symbols correspond to networks
  of $N=10^5$ nodes, and the lines represent the numerical
  simulations.}
\end{figure}

\section{Generalized attachment}
\label{sec:generalized}

The symmetry condition for the step-function $h(x)$ implies an
important result when part of the degree distribution (for example
around the head, i.e.~$k\to 0$) shows a power-law decay: The tail of
the distribution (i.e.~$k\to\infty$) also follows a power-law distribution,
but with a different exponent. To see this, let us assume that the
head of the distribution has the functional dependence $P(k) = \beta
k^{-\eta}$. If $\eta > 0$, this implies that the step-function
$h_l(x)$ for low degree nodes is given by $h_l(x) = {\beta}
k^{-\eta-1} / {1-\eta}$. By inverting this function, we get 
$$
x = \frac{\beta}{1-\eta} h_u(x)^{-\frac{1}{\eta-1}} ;
$$
and the distribution in the tail yields 
$$
P_u(k) = \frac{1}{1+\eta}\left( \frac{\beta}{1-\eta}\right)^{\frac{1}{\eta+1}} k^{-\eta_u},
$$ 
where $\eta_u = \eta/(\eta-1)$. In the limit $\eta
\to \infty$, (there is an exponential distribution for the head), it
implies $\eta_u \to 1$, i.e. we recover the previous result of
Eq.~(\ref{eq:basic-model-stationary-degree-distribution}). 
The
power-law distribution in the head and in the tail have the same
exponent when $\eta = 2$.

So far we have assumed that all nodes are selected at the same rate,
regardless of their position in the network. Depending on the context,
this assumption may not apply. In order to overcome this limitation, we
assume that nodes are selected at a rate which depends on their
position in the network. Note that the rate at which nodes are
selected affects only the frequency but not the way in which they
create or remove links. Therefore, the nestedness of the network is
preserved.  Moreover, in these nested structures, the nodes with the
same degree are indistinguishable, as only their degree rank in the
network is important.  We therefore assume that the node selection
rate $F$ is a function of the degree of the node.  As a simple
example, we set $F(k) = k^\eta + A$, where $A>0$ denotes the
idiosyncratic activity of every node, and $\eta>0$ a parameter governing
nonlinearly
the preferential selection of nodes with higher degree.  Using similar
arguments as in the derivation of
Eq.~(\ref{eq:cont:intermediate:full}), we can write the evolution of
the degree distribution as follows,
\begin{eqnarray}
  \partial_t P(k; t) &=& \frac{(1-2 \alpha) \eta}{\mathcal{N}} k^{\eta-1} P(k; t) \nonumber \\
   & &+ \frac{1-2 \alpha} {2\mathcal{N}}\left[ k^\eta+A \right] \partial_k P(k; t) \delta k + \mathcal{O}(\delta k^2). \nonumber
\end{eqnarray}
In the continuous limit, the stationary solution is given by 
$$
 P(k) = \frac{D}{A+k^{\eta}},
$$
where $D$ is a normalization constant such
that $\int_0^1 P(k) \, dk = 1$.  The solution reduces to the
exponential one when $\eta \to \infty$ and $A \ll N$.  In the general case,
the degree distribution exhibits two different power-law behaviors and
an inflection point. These two power-laws have the functional form
$P(k) \sim k^{-\eta}$ for the head of the distribution, and
consequently 
$$ 
P(k) \sim k^{-\frac{\eta}{\eta-1}}
$$ for the tail.
  
\section{Conclusions}
\label{sec:conclusions}

In this paper we have introduced a network formation process in which
link creation and removal is based on the position of the nodes in the
network measured by their centrality.  We have shown that the network
evolution is independent of the exact measure of centrality, and our results hold irrespective of whether degree centrality or any more general centrality measure that is based on walks or paths in the network is used. Thus, our model provides a general framework to study the evolution of networks under various measures of centrality. Moreover, we can show that the link formation decision of nodes does not require global information of the complete network structure. A further characteristic property of our model is that the emerging network structures are nested with a tunable degree of centralization, depending on the likelihood with which links are formed. This illustrates that both, a hierarchical organization as well as network nestedness can be the outcomes of a centrality based network formation process. Finally, we extend the model to allow for heterogeneous activity levels in the linking process of the nodes. We show that this generalization keeps the basic properties of the model unaltered, although the degree distribution is modified, and a restricted set of double power-law degree distributions is found. 

We have also discussed the broad range of applications of this kind of dynamics. In this context, it is worth mentioning the recent  empirical analysis of the European interbank payment network, which shows that our dynamic model matches closely the observed network pattern \citep{Cohen-Cole2011}.

\section*{ACKNOWLEDGEMENTS} 

MDK and CJT acknowledge financial support from SBF (Switzerland) and SNSF through research grants 100014\_126865, CR12I1\_125298 and PBEZP1\_131169 respectively. The authors thank F. Schweitzer for useful comments and discussion.

\bibliographystyle{apsrev4-1}
\bibliography{references-phys}

\end{document}